\documentclass[12pt]{spieman}  
\usepackage{amsmath,amsfonts,amssymb}
\usepackage{graphicx}
\usepackage{setspace}
\usepackage{tocloft}
\usepackage{caption}
\usepackage{soul}

\title{On-sky SiPM Performance Measurements for Millisecond to Sub-Microsecond Optical Source Variability Studies}
\author[a,*]{Albert Wai Kit Lau}
\author[b]{Mehdi Shafiee}
\author[b,c,d,e]{George F. Smoot}
\author[g,b]{Bruce Grossan}
\author[f]{Siyang Li}
\author[b]{Zhanat Maksut}
\affil[a]{Department of Physics, The Hong Kong University of Science and Technology}
\affil[b]{Energetic Cosmos Laboratory, Nazarbayev University, Kazakhstan}
\affil[c]{Institute for Advanced Study, Hong Kong University of Science and Technology}
\affil[d]{Lawrence Berkeley National Laboratory, Berkeley, USA}
\affil[e]{Université Sorbonne Paris Cité, Laboratoire APC-PCCP, Université Paris Diderot}
\affil[f]{Department of Physics, University of California, Berkeley, USA}
\affil[g]{Space Sciences Laboratory, University of California, Berkeley, USA}

\cftpagenumbersoff{figure}
\cftpagenumbersoff{table} 
\begin{document} 
\maketitle

\begin{abstract}
In our Ultra-Fast Astronomy (UFA) program, we aim to improve measurements of variability of astronomical targets on
millisecond and shorter time scales. In this work, we present initial on-sky measurements of the performance of silicon photomultiplier detectors (SiPMs) for UFA. 
We mounted two different SiPMs at the focal plane of the 0.7-meter aperture Nazarbayev University Transient Telescope at the Assy-Turgen Astrophysical Observatory (NUTTelA-TAO), with no filter in front of the detector. The $3mm\times3mm$ SiPM single-channel detectors have a field of view of $2.2716'\times2.2716'$. During the nights of 2019 October 28-29, 
we measured sky background, bright stars, and an artificial source with a 100Hz flashing frequency. We compared detected SiPM counts with Gaia satellite G-band flux values to show that our SiPMs have a linear response. With our two SiPMs (models S14520-3050VS and S14160-3050HS), we measured a dark current of $\sim$130 and $\sim$85 kilo counts per second (kcps),
and a sky background of $\sim$201 and $\sim$203 kcps, respectively. We measured an intrinsic crosstalk of 10.34$\%$ and 10.52$\%$ and derived a 5$\sigma$ sensitivity of 13.9 and 14.0 Gaia G-band magnitude for 200ms exposures, for the two detectors respectively. For a 10 $\mu$s window, and allowing a false alarm rate of once per 100 nights, we derived a sensitivity of 22 detected photons, or 6 Gaia G-band magnitudes. For nanosecond timescales, our detection is limited by crosstalk to 12 detected photons, which corresponds to a fluence of $\sim$155 photons per square meter.

\end{abstract}

\keywords{Ultra Fast Astronomy, Silicon photomultiplier, Millisecond optical variability}

{\noindent \footnotesize\textbf{*}Albert Wai Kit Lau,  \linkable{awklau@connect.ust.hk} }

\begin{spacing}{2}   
\section{Introduction}
\label{sect:intro}
 In our Ultra Fast Astronomy (UFA) program, we aim to survey the sky at sub-second (ns to ms) time scales in the optical-IR bands.
 \cite{10.1117/12.2548169} We already know of millisecond(ms) variability among the fastest-known sources of X-ray and optical variability and transients (e.g. X-ray binaries and pulsars). Fast radio bursts (FRBs) have ms variations in the radio frequencies; if there were accompanying emission in the optical \cite{hardy2017search}, it could be orders of magnitude faster. Occultation techniques with high time resolution can provide sub-milliarcsecond accuracy on direct measurement of stellar diameters \cite{richichi2014occultations, BenbowW2019Dmos}. Arguments have been made that optical bands are favorable for interstellar communication due to the high bandwidth, requiring short time scale sensitivity for optical SETI \cite{cosens2018panoramic}. In addition, we point out that probing shorter time scales at improved sensitivities could be a rich source of new discoveries, as this is a poorly covered part of the sensitivity-time scale parameter space. 
 
 The search for these sub-second transient events using traditional CCD-like detectors is limited by readout noise, readout rate and hence frame rate. For example, the HiPERCAM system, as one of the most sensitivity CCD-based fast optical camera, can provide $\sim 1$ frame per second (fps) on full readout mode or $\sim 1000$ fps on drift-mode \cite{dhillon2016hipercam}. Therefore, non-integrating photon detectors like Photomultiplier Tubes (PMTs), Silicon Photomultipliers (SiPMs) or more advanced superconductive photon counting sensors \cite{shafiee2019design, alzhanov2019cryogenic} are better suited for exploring even shorter time scales. Recently, we began measurements on the sky using a SiPM-based testing camera on the 0.7-m aperture Nazarbayev University Transient Telescope at the Assy-Turgen Astrophysical Observatory (NUTTelA-TAO)\cite{grossan2019emission}.  In this experiment, two different models of SiPM were used to measure sky background, bright stars and an artificial millisecond-period source. The experiment was performed on the nights of 2019 October 28 and 29, with an ambient temperature of roughly $0^{\circ}$C and a radiometric sky temperature (measured with a Boltwood Cloud Sensor II) in the range of $-25^{\circ}$C to $-30^{\circ}$C, with a slightly cloudy sky.
 
 In this paper we present our experimental setup, observation log and data processing method. 
In addition, we present our calculations of detection of lower limits at different time scales, along with our measurements of SiPM crosstalk, dark count and sky background. Finally, we propose some improvements to our system for future observations.
\section{Experimental setup}
\label{sect:setup}
Fig. \ref{fig:exp_setup} shows the setup for this experiment at Assy-Turgen Astrophysical Observatory. 

\begin{center}
    \includegraphics[width=\textwidth]{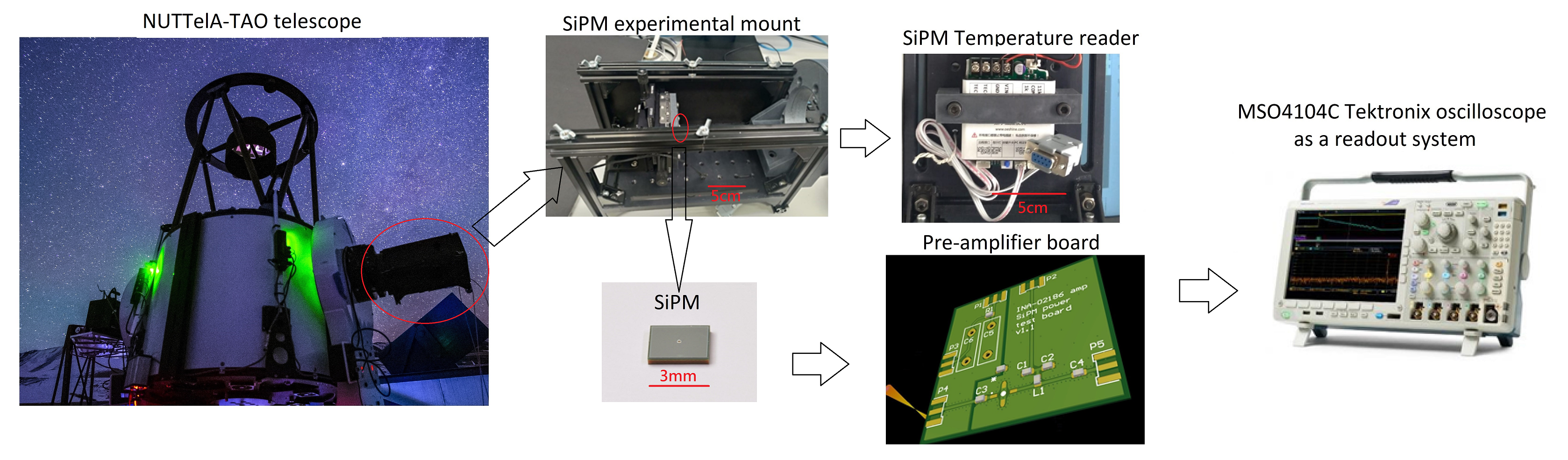}
    \captionof{figure}{Experimental Setup. The left image shows the telescope with the SiPM experiment mounted on the instrument port at right. The four center images show the SiPM mount, hardware and electronics. The far right image shows the oscilloscope used to view the signal and acquire the data.}
    \label{fig:exp_setup}
\end{center}

\subsection{SiPMs and Readout System}
\label{sect:system}
\subsubsection{SiPMs}
\label{sect:system:SiPM}
SiPMs are a relatively recent technology, giving photon counting from a solid-state chip. These detectors achieve single- and multiple-photon resolving power by massive parallel Geiger mode avalanche diodes (micropixels) which produce a photon gain \cite{saveliev2004recent} of $\sim10^6$. A micropixel can only receive one count within the recovery time, and multiple photons detection is achieved by summing up the signals from the parallel micropixels. Note that the micropixels, effectively summed internally, do not give spatial resolution within a single SiPM channel. \cite{mirzoyan2013sipm}

SiPMs have a high damage threshold for bright light exposure, a much lower working voltage and higher overall quantum efficiency compared to PMTs. Some trade-offs of SiPMs vs. PMTs include higher crosstalk and dark count noise \cite{buzhan2009cross}. In SiPMs, crosstalk is said to occur when one incoming photon triggers more than one micropixels; this occurs when nearby micropixels also contribute electrons due to electron leakage or re-emission in the semi-conductor. These are measured and discussed further in section 4.4 and section 5.

SiPMs' immunity to magnetic field and bright light damage provide easiness on mechanical and experimental design. In terms of signal integrity, SiPMs are also providing better signal amplitude linearity and time resolution compare to PMTs \cite{mirzoyan2013sipm}. As a more recent technology, the parameters of SiPM are still in rapid improvement. SiPM based Astronomy detectors like the next generation of imaging atmospheric Cherenkov telescopes are emerging. \cite{mirzoyan2013sipm, ambrosi2016large, buzhan2009cross} Therefore, we designed our fast astronomical detectors based on SiPM technology.

In this experiment, we used single channel SiPM models S14160-3050HS and S14520-3050VS from Hamamatsu \cite{yamamoto2019recent}. The manufacturer's specifications of these two models are given in Table \ref{tab:SiPM_spec} (at standard conditions at 25C$^{\circ}$, and standard overvoltage).

\begin{table}[hbt!]
\centering
\captionof{table}{Manufacturer's specifications of the SiPMs}
\begin{tabular}{ |c||c|c|}
\hline
Specifications & SiPM s14160-3050HS & SiPM s14520-3050VS\\
Testing voltage & breakdown voltage +2.70V & breakdown voltage +3.00V\\
\hline
Photosensitive area per channel & 3.0mm $\times$ 3.0mm & 3.0mm $\times$ 3.0mm\\
Micropixel pitch & 50$\mu m$& 50$\mu m$ \\
Micropixels per channel & 3531 & 3531\\
Spectral response range & Appendix \ref{appx:spectral} Fig. 13 &  Appendix \ref{appx:spectral} Fig. 14
\\
Peak photon detection efficiency & 50\% @ 450nm & 49\% @450nm \\
breakdown voltage & 38V & 38V\\
Temperature coefficient of breakdown voltage & 34mV/C$^{\circ}$&34mV/C$^{\circ}$\\
Standard overvoltage & 2.70V & 3.00V\\
Crosstalk & 7$\%$ & 5$\%$ \\
Dark count & 1Mcps & 600kcps \\
Photon gain & $2.5\times10^6$ & $2.8\times10^6$ \\
Optical Window & Silicone, 150$\mu m$ & Silicone, 150$\mu m$\\
Windows refractive index & 1.57 & 1.57\\
\hline
\multicolumn{3}{c}{}\\
\end{tabular}
\label{tab:SiPM_spec}
\end{table}

The dark count noise of SiPMs is significantly reduced at a lower temperature. During the experiment, the SiPMs ran at $\sim 0^{\circ}$C without any external cooling. Since the breakdown voltage of a SiPM also decreases with decreasing temperature, we carried out a breakdown voltage versus temperature measurement before the experiment at the observatory (appendix \ref{appx:calibration}). The SiPM voltage was maintained at the breakdown voltage plus the standard overvoltage throughout the experiment. The standard overvoltage values are provided by Hamamatsu: $2.7V$ for S14160-3050HS and $3V$ for S14520-3050VS . 

\subsubsection{Pre-amplifier circuit}
\label{sect:system:pre_amp}
The output signal of the SiPM is a charge pulse. We used a 50$\Omega$ shunt resistor to convert this signal to a voltage pulse, followed by a 30dB pre-amplifier to provide a larger amplitude signal for readout. The 30dB pre-amplifier design is based on the Monolithic Microwave Integrated Circuit (MMIC) low noise amplifier INA02186 \cite{hpamp} and gives a flat gain up to 1GHz with a simple circuit design. The SiPM is AC coupled to this pre-amplifier with each detected single photon yielding photoelectrons (P.E.) that produce a $\sim 5mV$ peak output.

\subsubsection{Data logging setup}
\label{sect:system:logging}
In the experiment, an MSO4104C Tektronix oscilloscope was used for data logging \cite{tektronix}. To simulate our proposed next-generation data logging scheme, we limited the sampling rate of the oscilloscope to $100$ megasamples per second ($100Msps$) with a bandwidth of $20MHz$, which is similar to common ADC data acquisition systems. The oscilloscope memory allowed us to take $2\times10^7$ data points for each measurement, corresponding to 0.2s observation time on each target under $100Msps$ sampling rate.

\subsection{Telescope specifications}
\label{sect:system:telescope}
The NUTTelA-TAO telescope (Fig.1) and instrumentation are described in (\cite{grossan2019emission}). The telescope has two Nasmyth focus ports, one occupied by the Burst Simultaneous Three-Channel Instrument (BSTI), so we mounted our instrument on the second port. The specifications of the telescope are given in table \ref{tab:tel_spec}. 

\begin{table}[h!]
\centering
\captionof{table}{NUTTelA-TAO Specifications}
\begin{tabular}{ |c||c|}
\hline
Diameter of primary mirror& 0.7m\\
Central obstruction	& 47\% of primary mirror diameter\\
Focal length & 4540mm (F6.5) \\
Effective light collection area & 2.998m$^2$\\
Optimal field of view & 70mm (0.86$^{\circ}$)\\
Image scale	 & 22$\mu m$ per arcsecond\\
\hline
\end{tabular}
\label{tab:tel_spec}
\end{table}

Our SiPMs have a single channel of 3mm$\times$3mm active area which views 2.2716'$\times$2.2716' at the NUTTelA-TAO Nasmyth focus under 4540mm focal length.

\section{Observation details}
\label{sect:obs}
We performed measurements on two consecutive nights, each with a different detector. Our list of observed targets and experimental conditions for each night are given below. 

\subsection{2019 October 28: Measurements with SiPM S14160-3050HS}
\label{sect:obs:obs_28}
On the first night of the test, the sky was slightly cloudy with a humidity of 60\% and a sky temperature of $-25^{\circ}$C. We chose to observe 
stars near zenith to minimize atmosphere effects. A dark portion of the sky (Dark Sky) with no source brighter than 18mag is also observed. 
Observations were made with SiPM S14160-3050HS, with sensor temperature of $-2.3^{\circ}$C to $-2.5^{\circ}$C. The observation log is presented in Table \ref{tab:obs_log_28}.

\begin{table}[h]
\centering
\captionof{table}{Observation log for 2019 October 28}
\label{sect:obs:obs_29}
\begin{tabular}{ |c|c||c|c|c|}
\hline
Target&Coordinate of pointing & Brightest star mag& Observe Time & Airmass\\
&(J2000)&(Gaia G-band)&(local: GMT+6)&\\
\hline
Star Field &RA 6h 51m 11.11s, DEC 58$^{\circ}$ 25' 2.8" & +7.93 & 04:24(+1d) & 1.05\\
Star Field &RA 6h 50m 54.42s, DEC 58$^{\circ}$ 23' 5.7" & +10.89 & 04:28(+1d) & 1.04\\
Star Field &RA 6h 56m 49.31s, DEC 58$^{\circ}$ 21' 36.5" & +13.11 & 04:32(+1d) & 1.04\\
Star Field &RA 6h 56m 40.19s, DEC 58$^{\circ}$ 28' 53.4" & +14.91 & 04:36(+1d) & 1.04\\
Dark Sky &RA 6h 56m 12.07s, DEC 58$^{\circ}$ 39' 26.2" & $>$+18 & 04:40(+1d) & 1.04\\
Dark test &dark count testing (shutter closed) & $\backslash$ & 04:45(+1d) & $\backslash$ \\
Star Field &RA 6h 53m 3s, DEC 59$^{\circ}$ 26' 53.8" & +5.2(saturated) & 04:56(+1d) & 1.04\\
\hline
\end{tabular}
\label{tab:obs_log_28}
\end{table}
\subsection{2019 October 29: Measurements with SiPM S14520-3050VS}
On the second night, we changed to SiPM S14520-3050VS. The sky was clearer, with a sky temperature of $-28^{\circ}$C.  The SiPM temperature was  $+2.4^{\circ}$C to $+2.6^{\circ}$C. We measured the same set of stars and dark sky as the previous night, but at greater airmass due to observing the field at a different time of night. The observation log is presented in table \ref{tab:obs_log_29}.
\begin{table}[h]
\centering
\captionof{table}{Observation log for 2019 October 29}
\begin{tabular}{ |c|c||c|c|c|}
\hline
Target& Coordinate of pointing  & Brightest star mag& Observe time & Airmass\\
&(J2000)&(Gaia G-band)&(local: GMT+6)&\\
\hline
Star Field &RA 6h 53m 3s, DEC 59$^{\circ}$ 26' 53.8" & +5.2(saturated) & 23:17 & 1.71\\
Star Field &RA 6h 51m 11.11s, DEC 58$^{\circ}$ 25' 2.8" & +7.93 & 23:21 & 1.66\\
Star Field &RA 6h 50m 54.42s, DEC 58$^{\circ}$ 23' 5.7" & +10.89 & 23:24 & 1.64\\
Star Field &RA 6h 56m 49.31s, DEC 58$^{\circ}$ 21' 36.5" & +13.11 & 23:27 & 1.66\\
Star Field &RA 6h 56m 40.19s, DEC 58$^{\circ}$ 28' 53.4" & +14.91 & 23:31 & 1.64\\
Dark Sky &RA 6h 56m 12.07s, DEC 58$^{\circ}$ 39' 26.2" & $>$+18 & 23:35 & 1.62\\
Dark test &dark count testing (shutter closed) & $\backslash$ & 23:39 & $\backslash$ \\
\hline
\end{tabular}
\label{tab:obs_log_29}
\end{table}

\section{Data processing and reduction}
\label{sect:datapro}
\subsection{Noise reduction and baseline cancellation}
\label{sect:datapro:noise}
The observatory power supply system produces frequent voltage spikes with frequencies $>10$MHz. Our pre-amplifier was powered from this source; the residual voltage spikes are the dominant source of noise in our data. We used a 21-point Savitzky–Golay low pass filter to reduce this noise while preserving the area under the data pulse, which varies with frequencies $<10$MHz. This filter limits our timing resolution to $\sim100ns$. 
Another noise source in our data is baseline drift, which comes from AC to DC power converters and is low frequency, mainly $<100$Hz. We apply a 200-point moving mean filter with high pass cut-off frequency of $500$kHz, and then subtract this smoothed baseline from the data to eliminate this drift.

The SiPM pulse has a sharp rise and a long decay tail of around a hundred ns. When multiple photons arrive within the decay (tail) time, pile-up occurs. To correct for pile-up, the start point of each pulse is adjusted to zero to ensure an accurate measurement. The top panel of Fig. \ref{fig:data_filter} shows part of the raw data measured using the S14520-3050VS SiPM and the bottom panel shows the processed data after noise reduction and baseline cancellation.

\begin{center}
    \includegraphics[width=\textwidth]{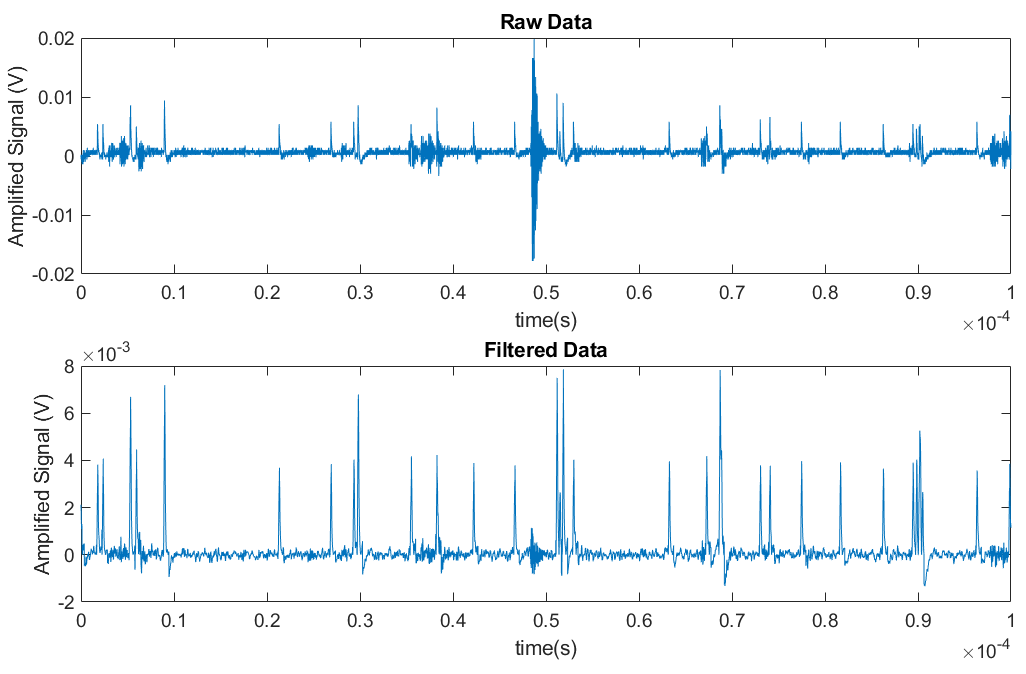}
    \captionof{figure}{(Top) Raw data, measured with S14520-3050VS, (bottom) Processed data, after filtration and baseline cancellation.}
    \label{fig:data_filter}
\end{center}

\subsection{Photoelectron identification}
\label{sect:datapro:pulse}
As mentioned before, SiPMs have a constant photon gain under constant overvoltage \cite{saveliev2004recent}. Therefore, the output charge (C = $\int I dt$) should give the number of arrival photons.

In our shunt and pre-amplifier electronics, the charge is converted to a voltage pulse, and ideally, the time integral of the pulse gives the number of detected photons over the integration. We therefore performed a trapezoid integration of each voltage pulse, and recorded the arrival time.

\label{sect:datapro:pe_peak}
Ideally, the pulse area histogram should be in discrete steps (as fractional photons are unphysical). In reality, the amount of calculated charge under each pulse can be affected by the electronic noise, imperfect pile-up correction, variation in the sensor, and readout system rounding errors. 
Each discrete pulse level then becomes broadened into a Gaussian distribution, and so the overall pulse area histogram becomes a sum of Gaussians.\cite{alvarez2013design, segreto2018liquid} To reduce these effects and correctly calculate the number of arrival photons, we implement a two-step Gaussian Mixture Model (GMM) with the steps below \cite{mclachlan2004finite}.
\begin{enumerate}
\item We fit a 2-peak GMM distribution to the data using an iterative Expectation-Maximization (EM) algorithm with a starting guess generated from a "k-means++" algorithm, which then generates a starting point by nearest mean clustering \cite{mclachlan2000peel, arthur2007k}. This algorithm provides us with the centroid and variance of the first and second photoelectron peaks (i.e., 1 P.E. and 2 P.E.).
\item From the centroid and variance of the first and second peaks, we estimate the centroid and variance of other P.E. peaks of the charge distribution using a linear regression forecast.
\item Having estimated values for the center and variance of each P.E. peak we perform GMM fitting again to obtain the precise number of arrival photons.
\end{enumerate}
We applied this second step of GMM fitting because the first auto GMM fitting did not provide precise information about peaks for $>$ 2 P.E.
\begin{center}
    \includegraphics[width=\textwidth]{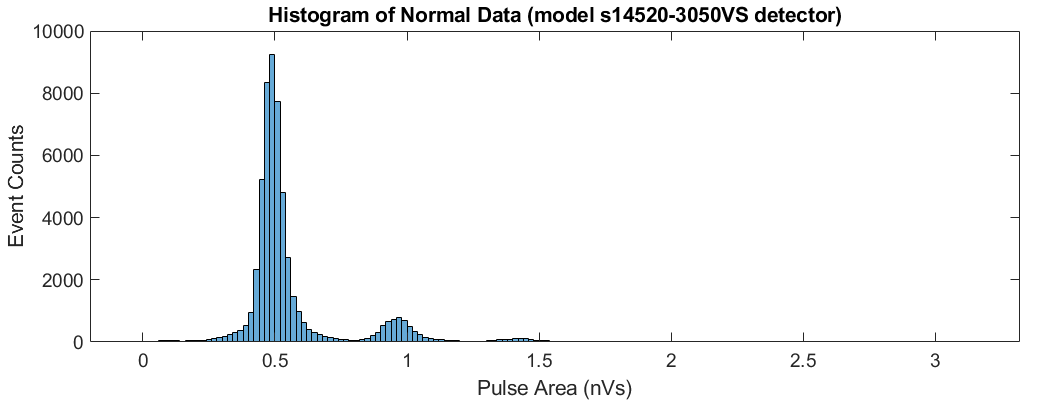}
    \captionof{figure}{A histogram plot of pulse area of different photon pulses. Notice the first and second peaks are both Gaussian-like, motivating the Gaussian Mixture Model used for fitting.}
    \label{fig:hist}
\end{center}

\subsection{Saturated data}
\label{sect:datapro:saturated}
Bright illumination of the SiPM causes saturation when the SiPM output pile-up occurs at a level such that single pulses can no longer be distinguished. In this case, the SiPM can still accept more photons, but we can not identify them on a photon counting basis. Fig. \ref{fig:sat} shows this case when we were observing 14 Lyn (a star with G-band magnitude of 5.2).
\begin{center}
    \includegraphics[width=\textwidth]{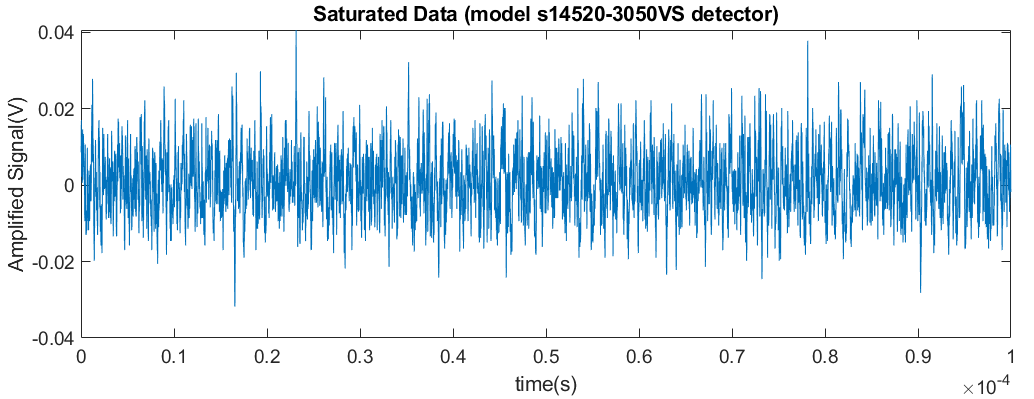}
    \captionof{figure}{Saturated data from measurement of the bright star 14 Lyn using SiPM S14520-3050VS.}
    \label{fig:sat}
\end{center}
From Fig.\ref{fig:sat_hist}, we can see that the GMM algorithm for photon number cannot be applied. By comparing Fig.\ref{fig:sat_hist} and Fig.\ref{fig:sat}, we find that the algorithm is not working well as 1 P.E and 2 P.E overlaps and they are not distinguishable from each other.

\begin{center}
    \includegraphics[width=\textwidth]{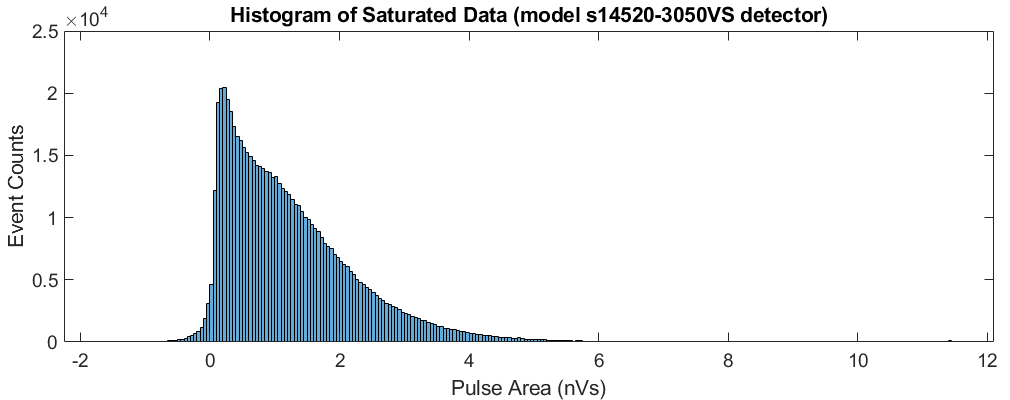}
    \captionof{figure}{Histogram plot of the saturated data from Fig. 4.}
    \label{fig:sat_hist}
\end{center}

\subsection{Crosstalk measurement}
\label{sect:datapro:crosstalk}
As mentioned above, when crosstalk occurs, one input photon will produce a 2 or higher P.E. pulse, which is indistinguishable from a real multiple P.E. signal \cite{collaboration2016description}.  Below, we find that crosstalk limits our ability to measure transient signals at very short time scales.  We therefore need to measure the crosstalk as a function of count rate for our devices.  

Measured crosstalk increases linearly with total count rate, with a zero-point, the "intrinsic crosstalk":
\begin{equation}
\begin{split}
    \text{Measured crosstalk rate} = &\text{intrinsic crosstalk $+$ multi-P.E. illumination crosstalk rate } \\
    &= \frac{\text{$>$ 1 P.E. SiPM count}}{\text{total count rate}}
\end{split}
\end{equation}
(We demonstrate this linearity in the next section.) To determine the crosstalk for any given  count rate, we analyze data with a range of multi-P.E. illumination and fit the intrinsic rate and slope of the function for each device. (Here we assume our illumination source is constant, so the number of multiple P.E. peaks ($>$ 1 P.E.) grows linearly with the flux (within expected fluctuations from Poisson statistics). The results of these fits are given below.

\section{Data Analysis}
\label{sect:analysis}
Our SiPMs is sensitive in the range of 300nm-900nm, similar but more blue-biased compared to the Gaia G-band spectral response. \cite{collaboration2016description,collaboration2018summary}. To flux calibrate our data, we therefore compared our results with the standard G-band flux data from the Gaia satellite archive \cite{collaboration2016description, collaboration2018summary}.
Our photometry analysis assumes integration over a circular aperture (diameter 2.563') of the same field of view as our detection field.  

\subsection{SiPM S14160-3050HS}
\label{sect:analysis:14160}
On the first night we used the SiPM S14160-3050HS  (See table \ref{tab:SiPM_spec}, \ref{tab:obs_log_28}). We corrected for airmass effects on our count rate for our photon flux calibration \cite{kasten1989revised}. Table \ref{tab:s14160_data_28} shows the calibrated flux, measured SiPM counts from stars, dark counts, and calculated crosstalk rate (saturated data are not listed).  

\begin{table}[h]
\centering
\captionof{table}{Measured data from SiPM S14160-3050HS}
\begin{tabular}{ |c|c||c|c|c|c|}
\hline
Target & Brightest star mag & calibrated flux& SiPM counts& crosstalk &Data\\
&(Gaia G-band)&(Gaia G-band, $e^-s^{-1}$) & in 200ms&(\%)&\\
\hline
Star Field & +7.93 & 12,563,646 & 341,726 & 23.2 & Fig.\ref{fig:s14160_HIP32890}\\
Star Field & +10.89 & 875,522 & 78,615 & 13.61& Fig.\ref{fig:s14160_10_89}\\
Star Field & +13.11 & 121,332 & 63,231 & 13.08 &Fig.\ref{fig:s14160_13_11}\\
Star Field & +14.91 & 26,911 & 58,415 & 12.82& Fig.\ref{fig:s14160_14_91} \\
Dark Sky & $>$+18 & 3,542 & 53,769 & 12.57&Fig.\ref{fig:s14160_empty} \\
Dark Test& $\backslash$ & $\backslash$ & 16,988 & 10.72 & Fig.\ref{fig:s14160_dark} \\
\hline
\end{tabular}
\label{tab:s14160_data_28}
\end{table}
Fig. 6 shows the airmass-corrected count rate versus the calibrated G-band flux.
\begin{center}
    \includegraphics[width=1\textwidth]{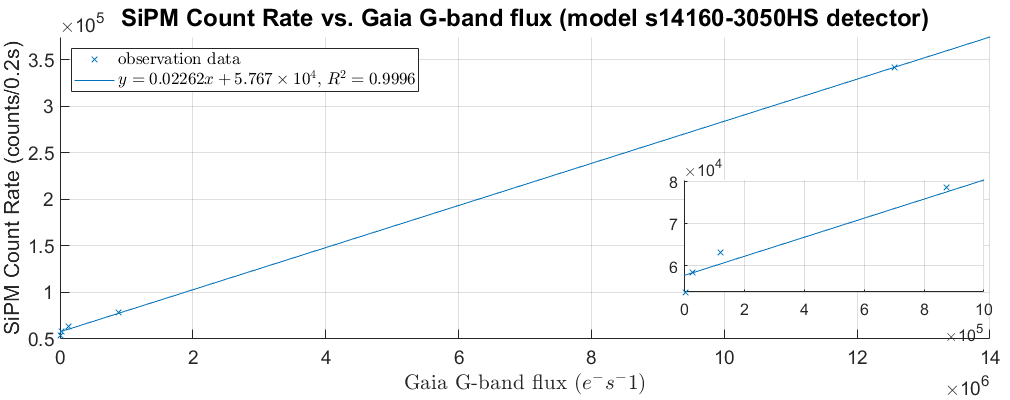}
    \captionof{figure}{SiPM airmass-corrected counts in 0.2 s versus calibrated G-band flux for SiPM s14160-3050HS. Inset gives data at low count rate.}
    \label{fig:s14160count}
\end{center}

The zero-flux intercept in Fig. \ref{fig:s14160count} gives dark counts plus sky background  within a 200ms window, 57,670 counts. We measured a SiPM dark count rate of $\sim$85 kilocounts per second (kcps; Table \ref{tab:s14160_data_28}). This gives a sky background of $\sim$203kcps. 

From the zero-flux intercept, and its uncertainty $\sigma_{intercept} = \sqrt{57670} = 240$ counts in 200 ms, we can calculte the $5\sigma$ sensitivity to be 1200 counts, corresponding to a Gaia G-band flux of 53,050 $e^-s^{-1}$, equivalent to a 13.9 G-band mag point source (assuming no atmospheric extinction). If we increase the integration time to 1 second, the SNR scales $\sqrt{t}$, giving a $5\sigma$ detection limit of 14.7 G-band mag in 1 s. 
 
 Fig. \ref{fig:s14160crosstalk} shows the fit of measured crosstalk versus count rate. We measured the intrinsic or zero-point crosstalk of SiPM S14160-3050HS to be 10.52$\%$.
 \begin{center}
    \includegraphics[width=1\textwidth]{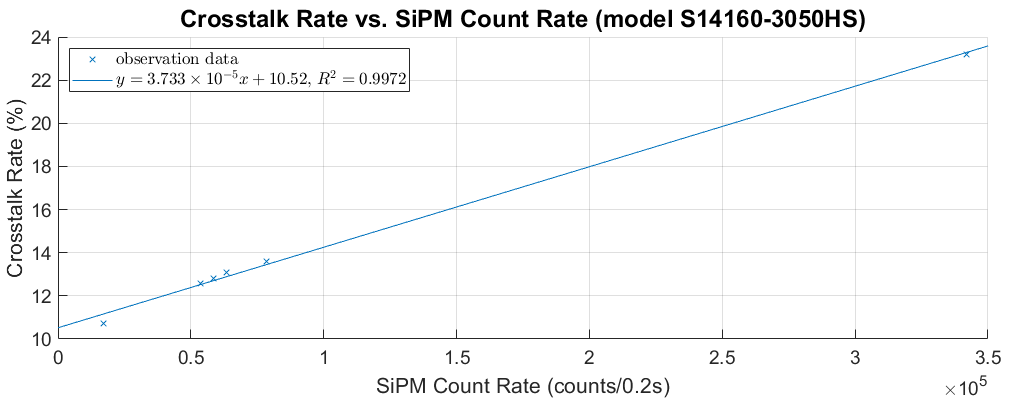}
    \captionof{figure}{Fit of measured crosstalk versus count rate for SiPM s14160-3050HS}
    \label{fig:s14160crosstalk}
\end{center}

\subsection{SiPM S14520-3050VS}
\label{sect:analysis:14520}
On the second night we used SiPM S14520-3050VS (See Table \ref{tab:SiPM_spec}, \ref{tab:obs_log_29}). We carried out the same measurements as on the first night, 
with results given in Table \ref{tab:s14520_data_29}.

\begin{table}[h]
\centering
\captionof{table}{Measured data from SiPM S14520-3050VS}
\begin{tabular}{ |c|c||c|c|c|c|}
\hline
Target & Brightest star mag & calibrated flux& SiPM counts& crosstalk &Data\\
&(Gaia G-band)&(Gaia G-band, $e^-s^{-1}$) & in 200ms&(\%)&\\
\hline
Star Field & +7.93  & 7,946,884 & 304,481 & 23.34 &Fig.\ref{fig:s14520_HIP32890}\\
Star Field & +10.89 & 555,209 & 82,317  & 13.94&Fig.\ref{fig:s14520_10_89}\\
Star Field & +13.11 & 76,015 & 69,858 & 13.27&Fig.\ref{fig:s14520_13_11}\\
Star Field & +14.91 & 17,065 & 67,854 & 12.9&Fig.\ref{fig:s14520_14_91}\\
Dark Sky & $>$+18 & 2,274 & 64,171 & 13.11&Fig.\ref{fig:s14520_empty}\\
Dark Test& $\backslash$ & $\backslash$ & 26,103 & 11.65 &Fig.\ref{fig:s14520_dark}\\
\hline
\end{tabular}
\label{tab:s14520_data_29}
\end{table}

Fig. \ref{fig:s14520count} shows SiPM count rate versus calibrated G-band flux.
\begin{center}
    \includegraphics[width=\textwidth]{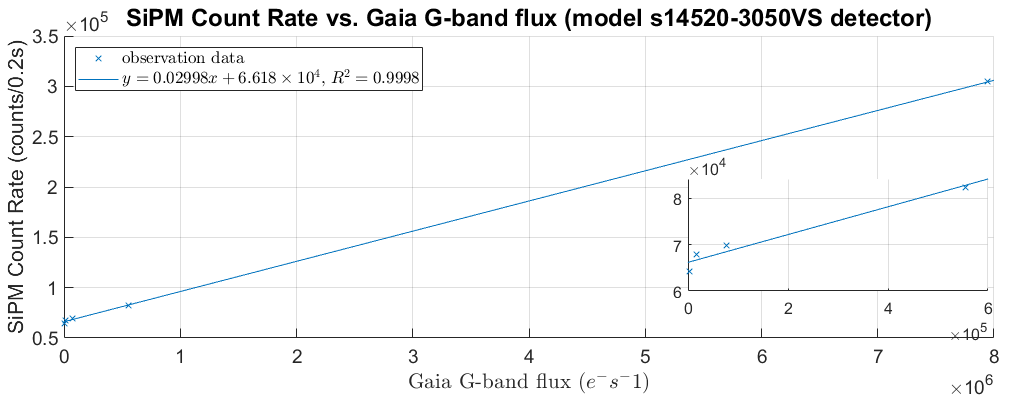}
    \captionof{figure}{Count rate versus calibrated G-band flux for SiPM S14520-3050VS. Inset gives data at low count rate.}
    \label{fig:s14520count}
\end{center}

From Fig. \ref{fig:s14520count}, and Table \ref{tab:s14520_data_29}, the SiPM dark count rate is $\sim$130kcps while the sky background is $\sim$201kcps (after subtracting the dark counts). As above, from $\sigma_{intercept} = 248$ counts, we find the $5\sigma$ sensitivity to be 1240 counts within 200ms, corresponding to to a Gaia G-band flux of $41,361e^-s^{-1}$, or a 14.0 G-band mag point source detected in 200ms (assuming no atmospheric extinction). Scaling by $t^{-1/2}$ gives a detection limit of 14.9 G-band mag in 1 s. 
 
Fig. \ref{fig:s14520crosstalk} shows the fit of the measured crosstalk versus count rate, giving an intrinsic crosstalk for SiPM S14520-3050VS of 10.34$\%$.
\begin{center}
    \includegraphics[width=\textwidth]{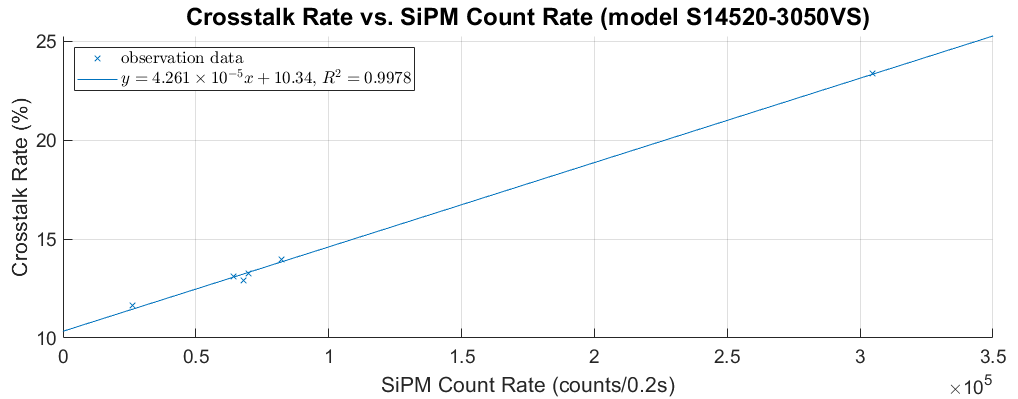}
    \captionof{figure}{Measured crosstalk versus count rate for SiPM S14520-3050VS.}
    \label{fig:s14520crosstalk}
\end{center}

\subsection{Time domain analysis}
\label{sect:analysis:time}
A prime motivation for our use of a SiPM detector is the ability to resolve events at short time scales. During all our observations of stars, no significant change in brightness was measured, as expected. We applied both $10\mu s$ and $1ms$ moving mean filters to produce smoothed light curves for all observations. These light curves, together with the histogram plot of photon statistics and corresponding Gaia star chart of each observation, are attached in the appendix \ref{appx:data_14160} and \ref{appx:data_14520}. 
\subsubsection{Synthetic light source test}
\label{sect:analysis:time:synthetic}
To simulate the response of our detectors to a fast transient signal, we scattered an artificial light source with a $\sim 100Hz$ square wave modulation into the telescope in the middle of our observations.
\begin{center}
    \includegraphics[width=1\textwidth]{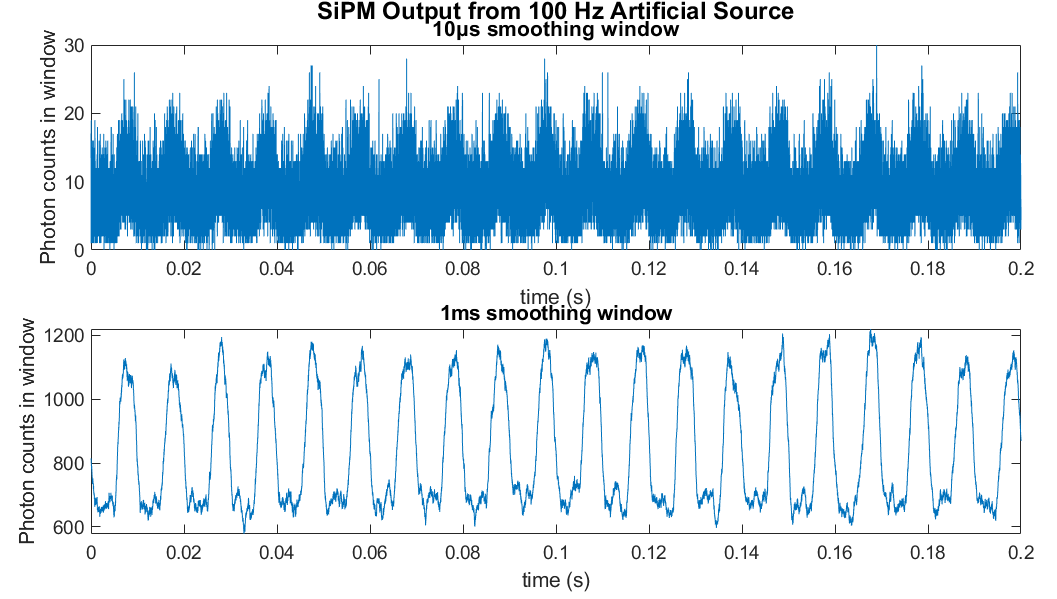}
    \captionof{figure}{10$\mu$s (top) and 1ms (bottom) smoothed signal from artificial 100Hz scattered square wave light.}
    \label{fig:100Hz}
\end{center}
In Fig. \ref{fig:100Hz}, the 100Hz signal data are shown using moving-mean filters with $10\mu s$ and $1ms$ windows. The $1ms$ window gives much better SNR compared to $10\mu s$ as expected. The spectrum of the light source is attached in Fig. \ref{fig:artificial_spec}, measured by a AvaSpec-2048 spectrometer with background removal.

\begin{center}
    \includegraphics[width=0.8\textwidth]{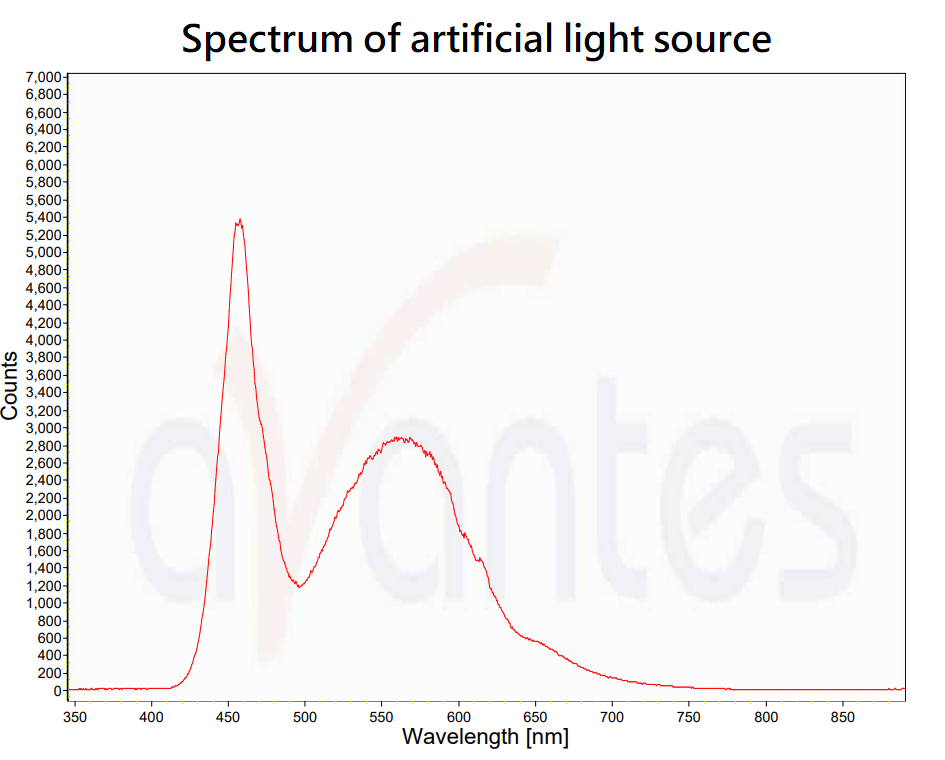}
    \captionof{figure}{spectrum of the 100Hz artificial light source used.}
    \label{fig:artificial_spec}
\end{center}

\subsubsection{Sensitivity vs time scale}
\label{sect:analysis:time:sensitivity}
Here we derive our detection limits for fast transients at various time scales. If we consider a dark portion of the sky, what is the minimum brightness of transient that we can detect?

Consider the measured dark sky using detector model S14520-3060VS, as given in table \ref{tab:obs_log_29} and \ref{tab:s14520_data_29}, we measured 64,171 photons within 200ms.

For fast transients, we measure only small numbers of photons and therefore need to consider shot noise (Poisson fluctuations); we are not in the Gaussian regime. We therefore fit the data to a Poisson distribution:
\begin{center}
    \includegraphics[width=1\textwidth]{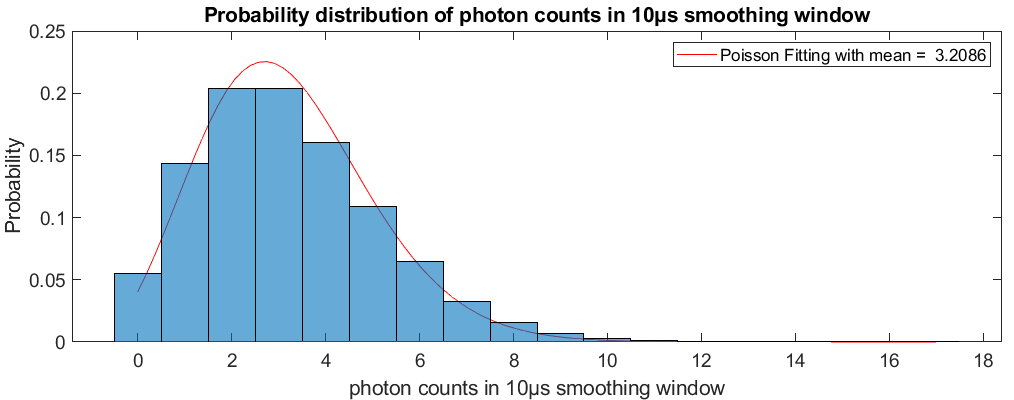}
    \captionof{figure}{Poisson fit to SiPM measurement of  dark sky}
    \label{fig:10us_poisson_fit}
\end{center}

From Fig. \ref{fig:10us_poisson_fit}, for SiPM S14520-3050VS, we find that the mean photon count rate is 3.2086 per 10$\mu s$. If we select a desired false alarm rate of less than 1 event per 100 night ($\sim 5 \times 10^{6} s$), we would then need a false trigger rate of less than 1 per $5\times 10^{11}$ time bins. From simple Poisson statistics, this corresponds to a detection threshold of 22 photons detected within 10$\mu s$. Comparing to the above count rate versus calibrated G-band flux (Fig. \ref{fig:s14520count}), the transient should give a G-band flux of at least $\sim 7.6 \times 10^{7} e^-s^{-1}$ ($\sim$6 G-band mag) in the 10$\mu s$ bin. Even for a false alarm rate as high as 1 event per night, we still need 20 photons detected within 10$\mu s$, or $\sim6.9 \times 10^{7} e^-s^{-1}$ ($\sim$6.1 G-band mag) in the 10$\mu s$ bin. 

Consider nanosecond time scale transients and a false alarm rate $<<$ 1 false alarm per 100 nights. From the same dark sky observation, we have 64,171 photons detected in 200ms (table \ref{tab:obs_log_29} and \ref{tab:s14520_data_29}) which corresponds to $~1.5\times 10^{12}$ photons per 100 nights. In nanosecond time bins, photons will pile up, and be detected as a multi-P.E. event. Such events have a low occurrence rate in the absence of crosstalk. For example, we got 64,171 photons detected in 200ms, correspond to one photon per $\sim3\mu s$. If we consider the SiPM decay(tail) time ofjbjh 100ns, then the natural rate of concurrent photons received is just $100ns/3\mu s = 3\%$. However, the measured crosstalk (which counts also the natural photons pile up) is $13.11\%$, much higher than $3\%$. This limits our sensitivity for nanosecond time scale transients. 

To maintain one false alarm per 100 nights, less than one event should go above threshold among all $~1.5\times 10^{12}$ photons detected, i.e. trigger probability less than $1/1.5\times 10^{12}$. With measured crosstalk of $13.11\%$, we should require $(13.11\%)^N \leq 1/1.5\times 10^{12}$, where N is the trigger threshold in number of photons. Solving gives a trigger threshold of least 14 P.E.. Considering an average SiPM detection efficiency of $\sim 30\%$ over 300nm to 900nm range and collection area $0.2998m^2$ of the NUTTelA-TAO system, this gives a fluence threshold of 155 photons per square meter within nanoseconds.

\section{Future improvements and plans}
\label{sect:improve}
For either model of SiPM, we can detect stars brighter than $\sim$14 mag in 200 ms bins, and saturation occurs at $\sim$7 mag. To perform astrophysical measurements for fast transient events, we need to have a higher dynamic range. For example, the Crab pulsar $PSR B0531+21$  itself is a 16.5 V-band magnitude source embedded in an 8.4 V-band magnitude nebula \cite{kinkhabwala2000multifrequency}. Therefore, we need to improve our ability to measure a faint point source against a bright nearby background. 

Currently, the SiPM sensitivity and dynamic range are mainly limited by spatial resolution. 
Our sky background is $\sim$200kcps for both detectors. A sensor with 
smaller physical size would have a significantly lower background rate. If the sensor were to view 10"$\times$10", the equivalent sky noise would be reduced by 185 times, a roughly 5.7 magnitude improvement. The same improvement would be achieved in dark count rate, since this scales with sensor area. Therefore, we aim to have a much smaller SiPM-based detector in our next iteration.

Both SiPMs have higher intrinsic crosstalk than the manufacturer’s specifications (10.52\% versus 7\% for S14160-3050HS and 10.34\% versus 5\% for S14520-3050VS). Possible causes include internal reflections from mounts and telescope mirrors, or a thin layer of frozen moisture on the sensor surface which we discovered on daytime inspection after the observation nights. We intend to pursue reduction of these possible causes. 

In order to improve our readout system, we are in the process of developing a standalone FPGA-based system\cite{shafiee2016analysis, shafiee2017experimental}. We are planning to use a system-on-chip (SoC) FPGA with an integrated ARM core to host a linux-based system. The SiPM signal should be converted to digital data using a high speed ADC, and then logged by the FPGA. The FPGA can then calculate the statistical significance of putative transients in real-time. If any candidate event is detected, the system could record the raw photon arrival information and alert the observer automatically to enable additional follow-up.  

During this experiment, we observed strong noise from the observatory power supply. To remove the noise, we will develop an isolated power supply for the detectors and the analog readout system. 

\section{Conclusion}
\label{sect:conclusion}
Initial measurements of two SIPM detectors for Ultra-Fast Astronomy have been successfully performed on the NUTTelA-TAO telescope. The 3mm $\times$ 3mm SiPM sensors provide a detection limit of, respectively, 13.9 (S14160-3050HS) and 14.0 G-band mag (S14520-3050VS) in a 200ms time window. The sky background and intrinsic crosstalk were measured to be $\sim$201, 203 kcps and $10.34\%$, $10.52\%$ for the S14160-3050HS and S14520-3050VS model detectors, respectively. For a false alarm rate of once per 100 nights, the 10$\mu s$ detection limit is 22 detected photons, which corresponds to a 6 G-band mag fast transient. At the shortest timescales, we derive a limiting fluence of $\sim$155 photons per square meter within 100ns, limited by crosstalk. The crosstalk is higher than expected, possibly due to internal reflections or defects in our detectors. We plan to improve our system resolution by reducing the detector size, which reduces sky background and dark counts. Better astrophysical measurements should also be achieved by crosstalk and power supply noise reduction, readout electronics improvement to handle longer data streams. These changes will allow us to search for transients on short timescales in astrophysical applications which is the main goal of our UFA program. 

\appendix
\newpage
\section{SiPM models spectral response and Gaia G-band transmittance}
\label{appx:spectral}
\begin{center}
    \includegraphics[width=0.6\textwidth]{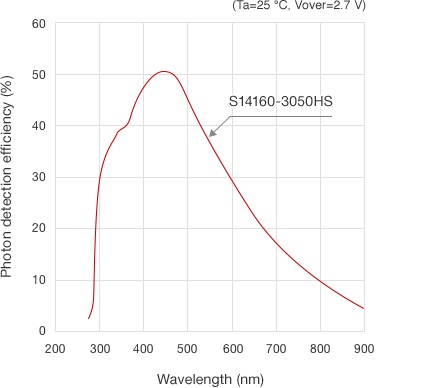}
    \captionof{figure}{Hamamatsu SiPM S14160-3050HS spectral response curve}
    \label{fig:s14160-spec}
\end{center}
\begin{center}
    \includegraphics[width=0.6\textwidth]{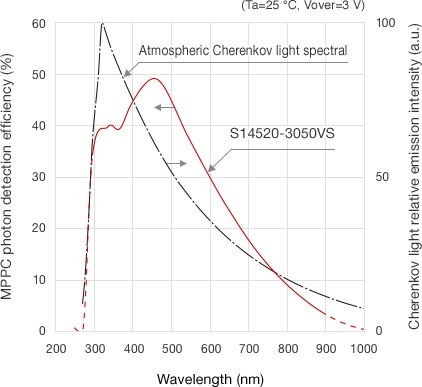}
    \captionof{figure}{Hamamatsu SiPM S14520-3050VS spectral response curve}
    \label{fig:s14520-spec}
\end{center}
Credit to Hamamatsu photonics
\begin{center}
    \includegraphics[width=0.6\textwidth]{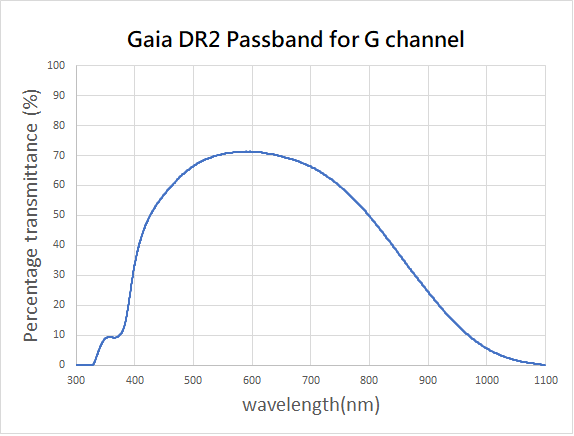}
    \captionof{figure}{Passband of Gaia DR2 G-band}
    \label{fig:gaia_g}
\end{center}
Data From Gaia collaboration.\cite{collaboration2018summary, collaboration2016description}

\newpage
\section{SiPM units breakdown voltage versus temperature calibration curves, tested in laboratory condition}
\label{appx:calibration}
\begin{center}
    \includegraphics[width=\textwidth]{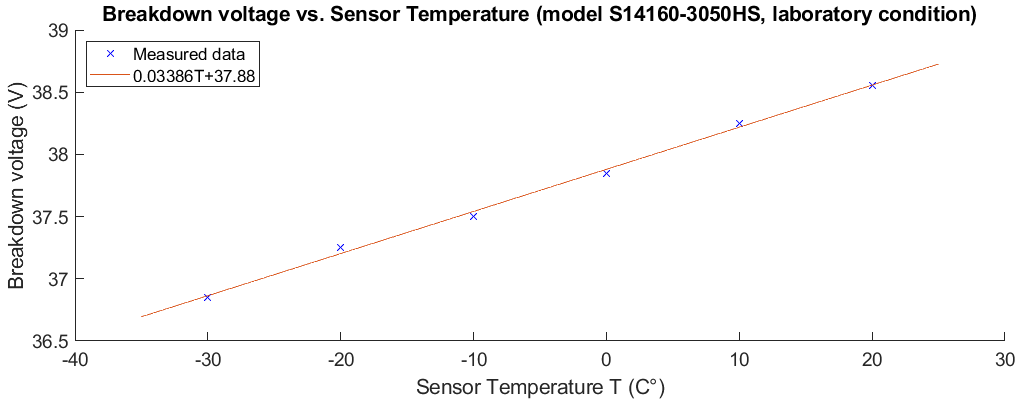}
    \captionof{figure}{Hamamatsu SiPM S14160-3050HS breakdown voltage versus temperature calibration curve}
    \label{fig:s14160-TV}
\end{center}
\begin{center}
    \includegraphics[width=\textwidth]{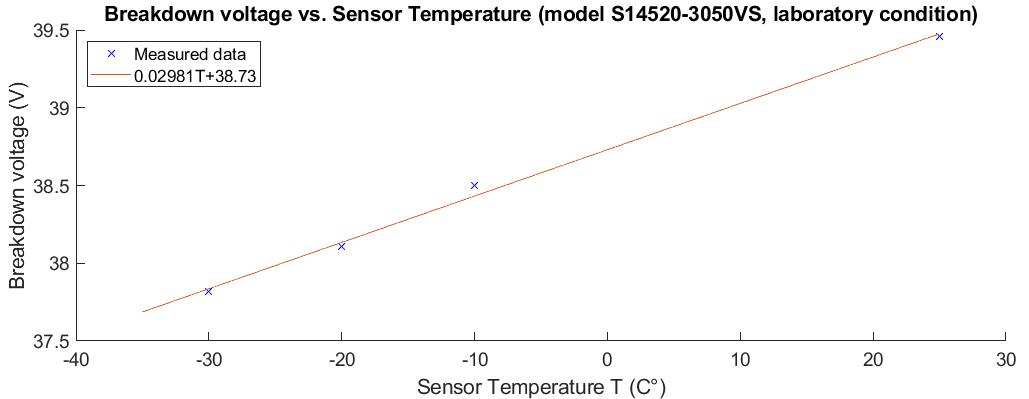}
    \captionof{figure}{Hamamatsu SiPM S14520-3050VS breakdown voltage versus temperature calibration curve}
    \label{fig:s14520-TV}
\end{center}

\newpage
\section{Analysed Data from S14160-3050HS}
\label{appx:data_14160}
\begin{center}
    \includegraphics[width=1.1\textwidth]{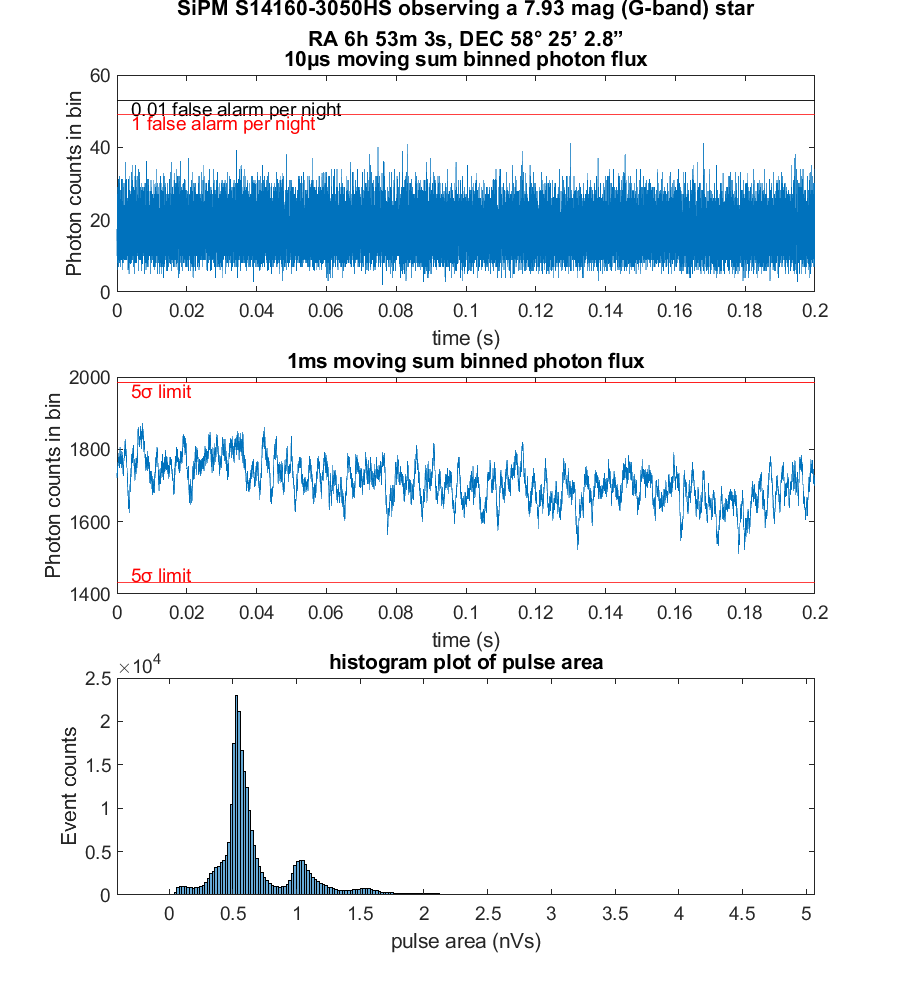}
    \captionof{figure}{SiPM S14160-3050HS observing HIP32890}
    \label{fig:s14160_HIP32890}
\end{center}
\begin{center}
    \includegraphics[width=1.1\textwidth]{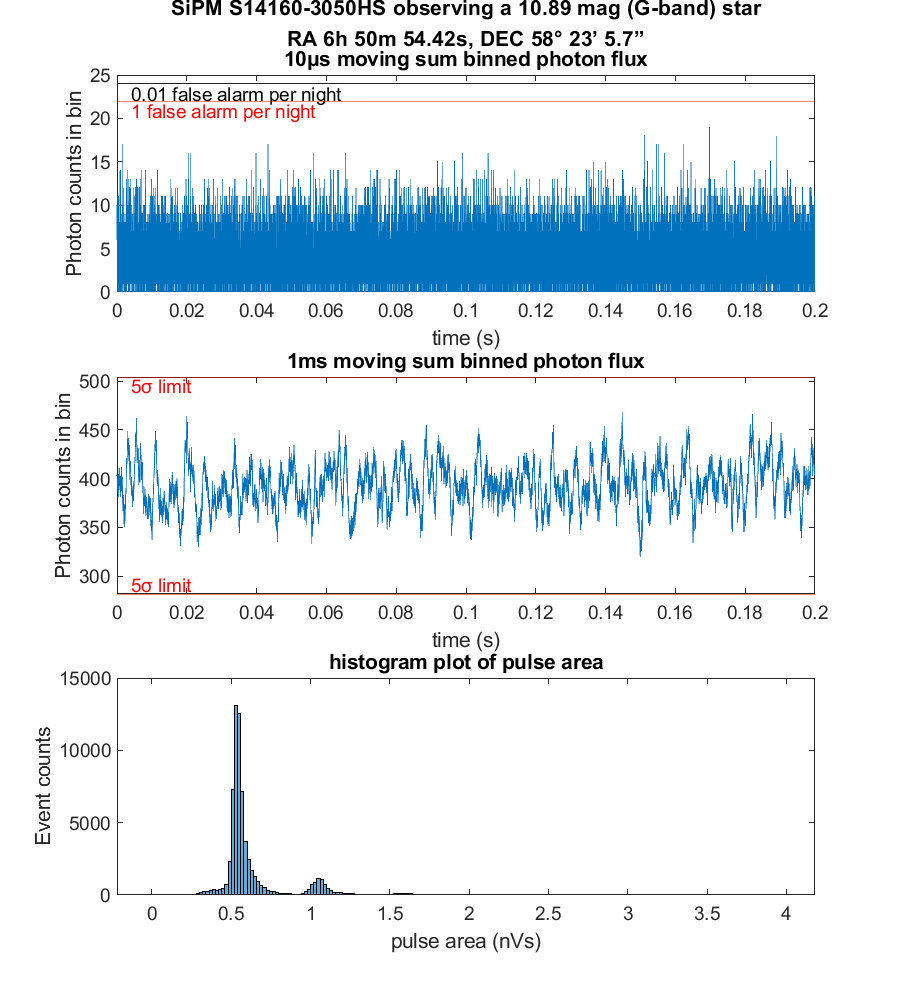}
    \captionof{figure}{SiPM S14160-3050HS observing a 10.89 mag (G-band) star}
    \label{fig:s14160_10_89}
\end{center}
\begin{center}
    \includegraphics[width=1.1\textwidth]{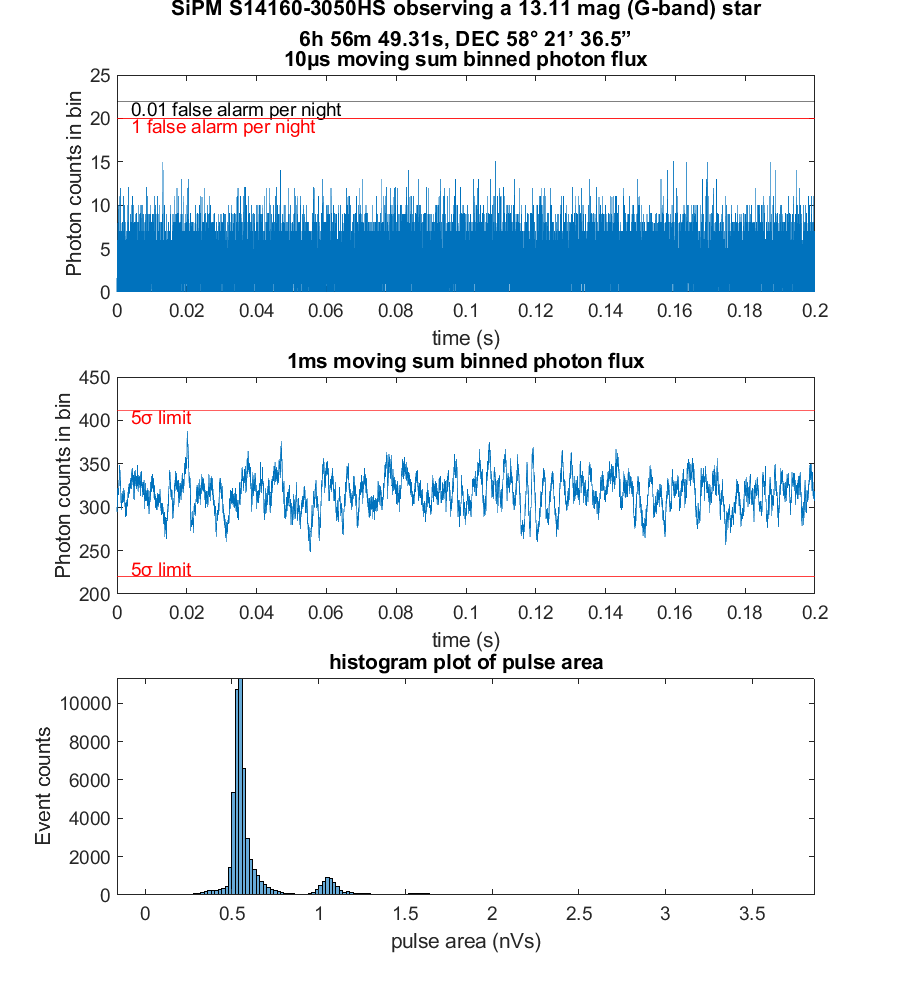}
    \captionof{figure}{SiPM S14160-3050HS observing a 13.11 mag (G-band) star}
    \label{fig:s14160_13_11}
\end{center}
\begin{center}
    \includegraphics[width=1.1\textwidth]{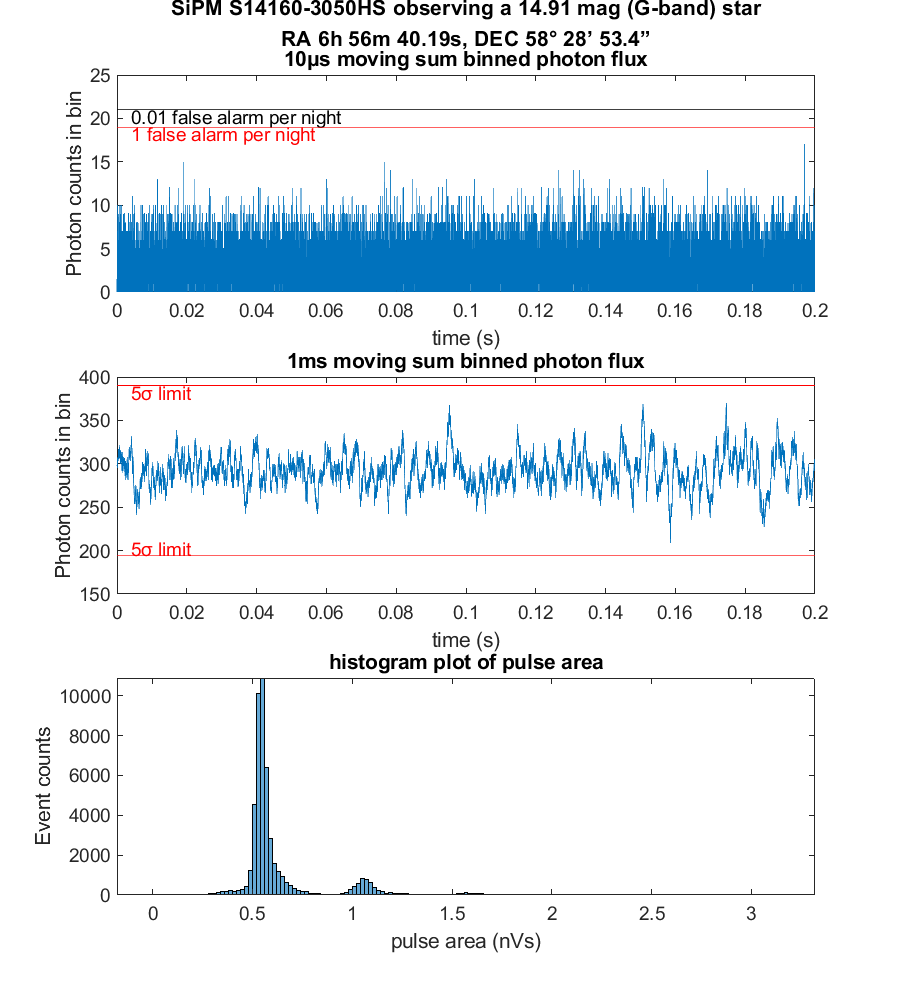}
    \captionof{figure}{SiPM S14160-3050HS observing a 14.91 mag (G-band) star}
    \label{fig:s14160_14_91}
\end{center}
\begin{center}
    \includegraphics[width=1.1\textwidth]{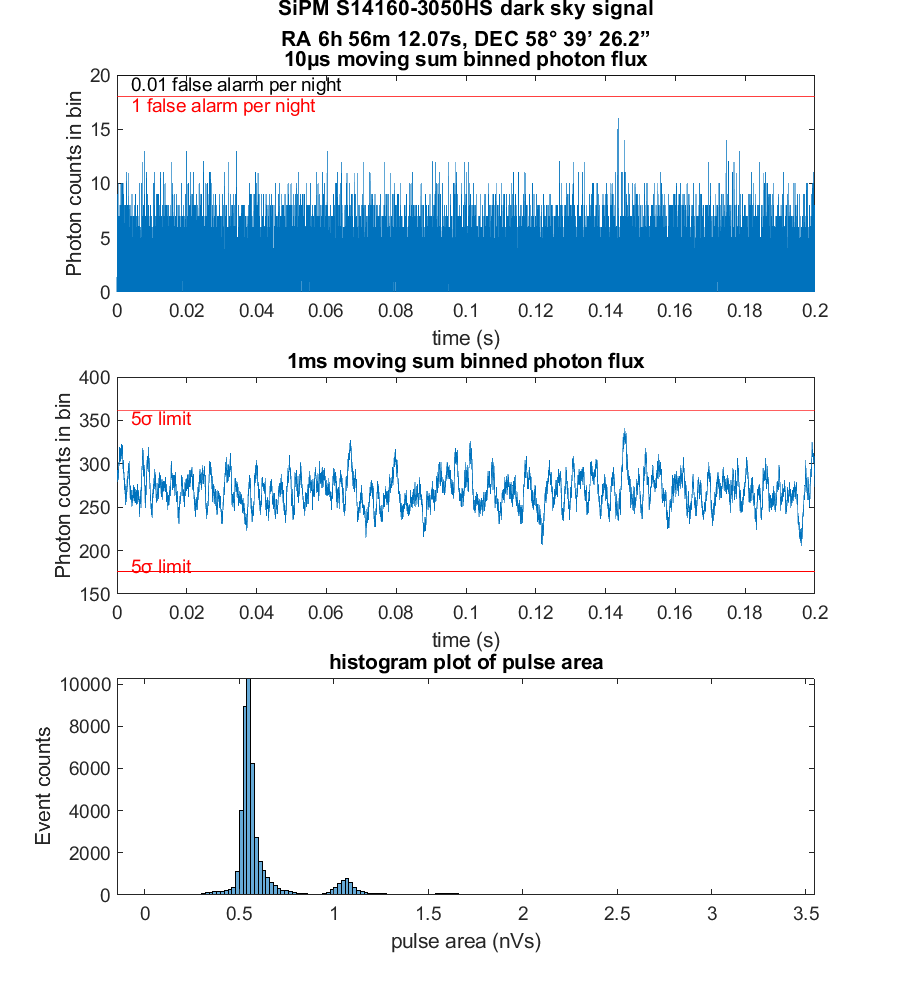}
    \captionof{figure}{SiPM S14160-3050HS observing area without star brighter than 18 mag (G-band)}
    \label{fig:s14160_empty}
\end{center}
\begin{center}
    \includegraphics[width=1.1\textwidth]{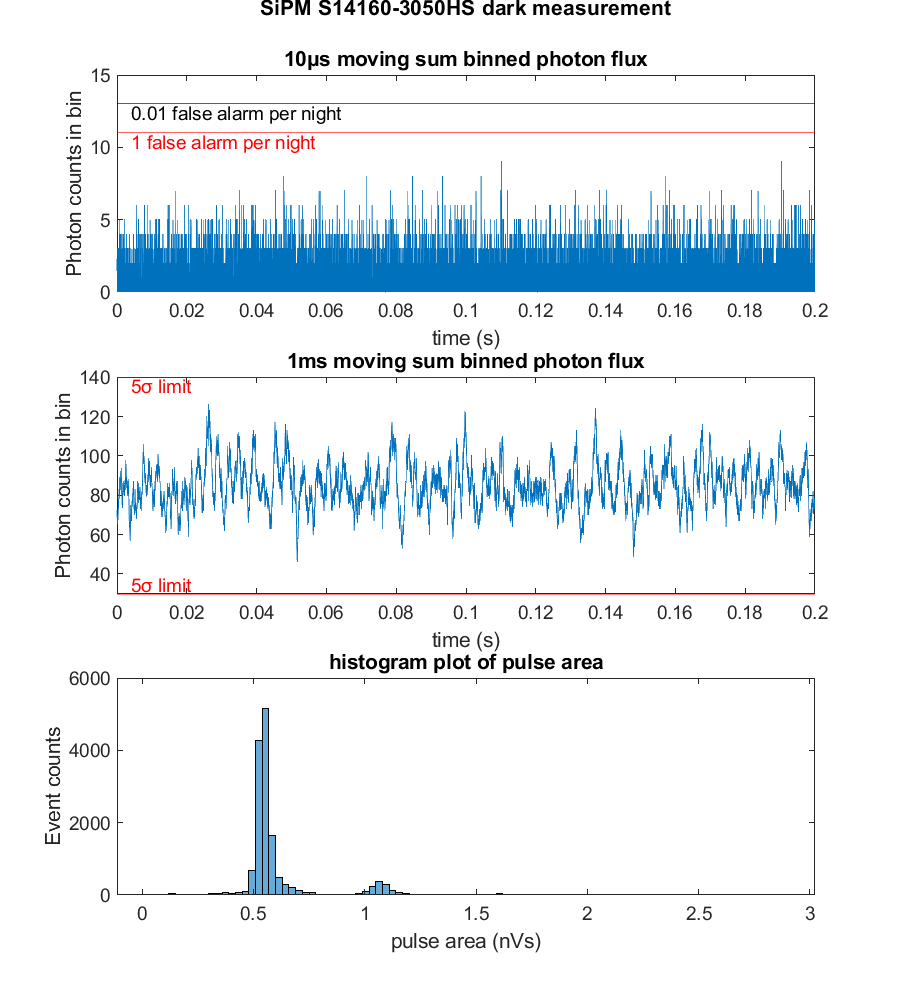}
    \captionof{figure}{SiPM S14160-3050HS dark measurement on site}
    \label{fig:s14160_dark}
\end{center}

\section{Analysed Data from S14520-3050VS}
\label{appx:data_14520}
\begin{center}
    \includegraphics[width=1.1\textwidth]{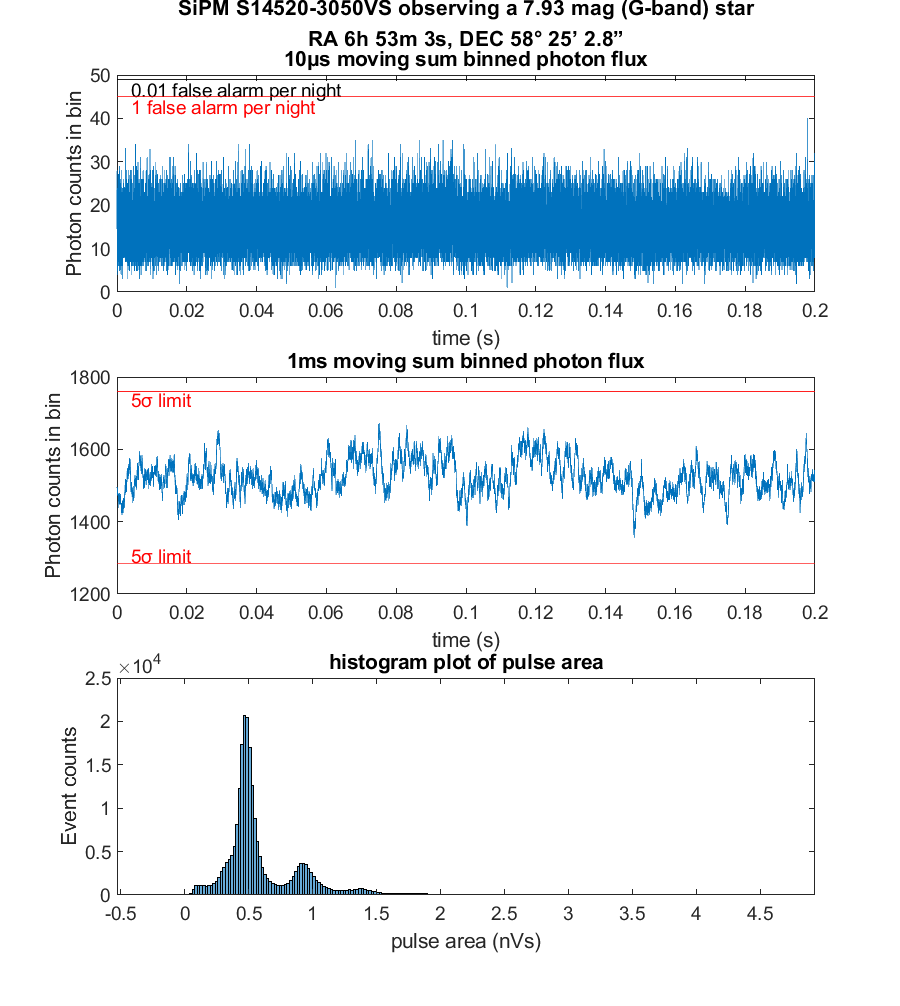}
    \captionof{figure}{SiPM S14520-3050HS observing HIP32890}
    \label{fig:s14520_HIP32890}
\end{center}
\begin{center}
    \includegraphics[width=1.1\textwidth]{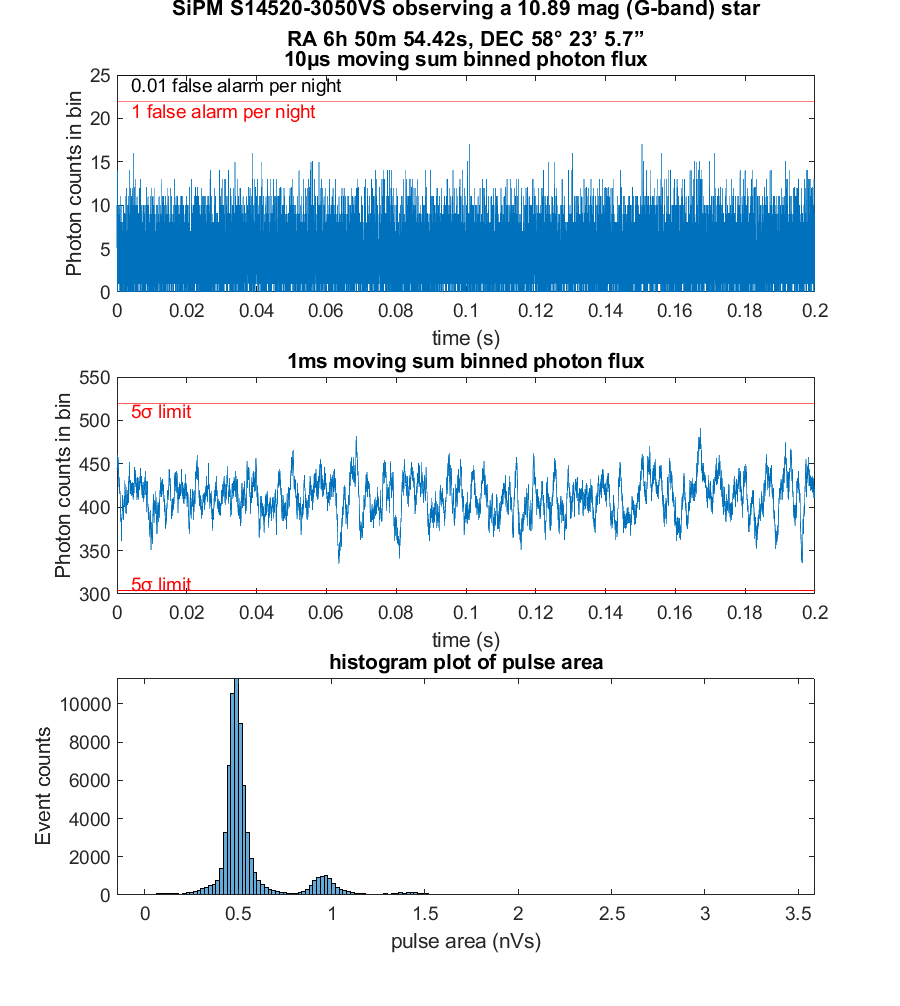}
    \captionof{figure}{SiPM S14520-3050HS observing a 10.89 mag (G-band) star}
    \label{fig:s14520_10_89}
\end{center}
\begin{center}
    \includegraphics[width=1.1\textwidth]{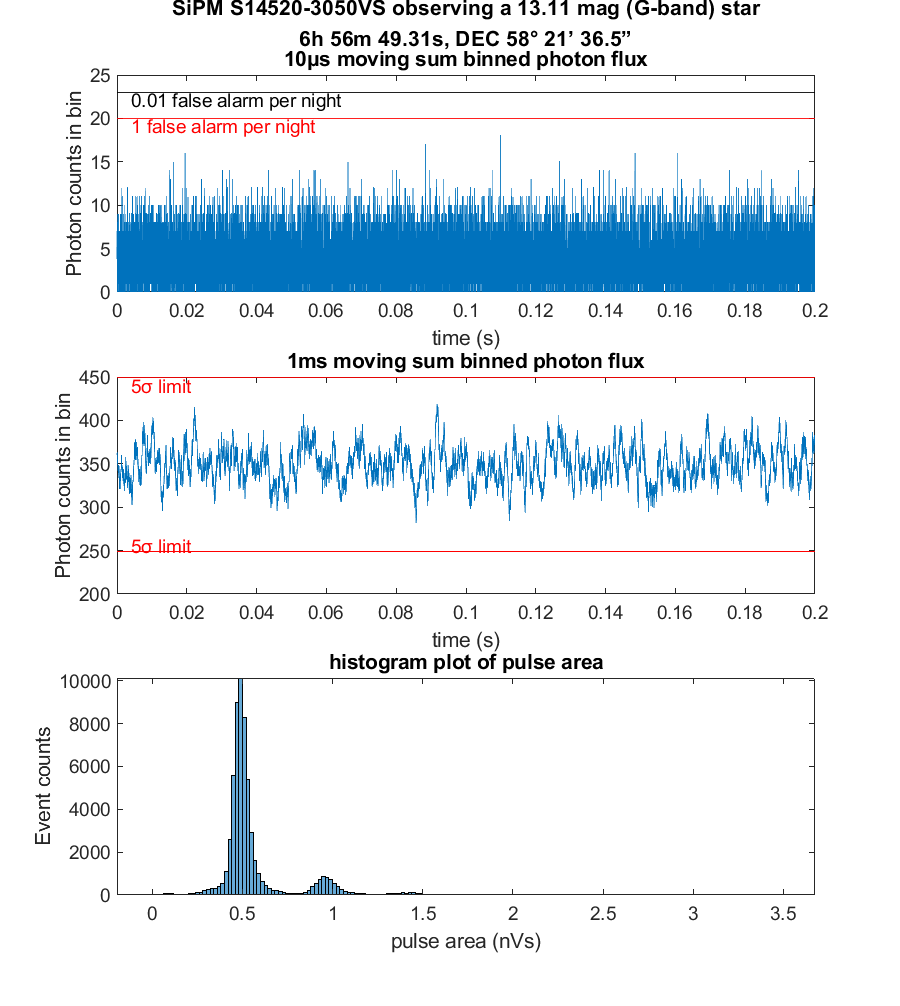}
    \captionof{figure}{SiPM S14520-3050HS observing a 13.11 mag (G-band) star}
    \label{fig:s14520_13_11}
\end{center}
\begin{center}
    \includegraphics[width=1.1\textwidth]{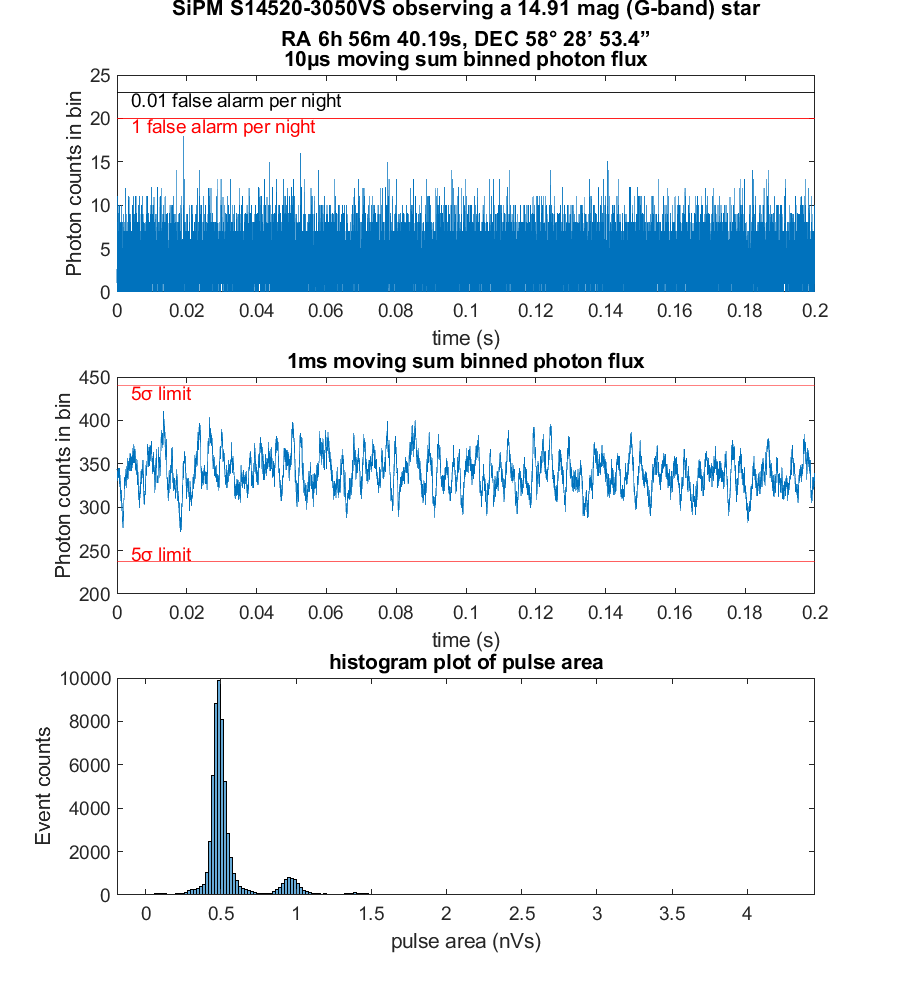}
    \captionof{figure}{SiPM S14520-3050HS observing a 14.91 mag (G-band) star}
    \label{fig:s14520_14_91}
\end{center}
\begin{center}
    \includegraphics[width=1.1\textwidth]{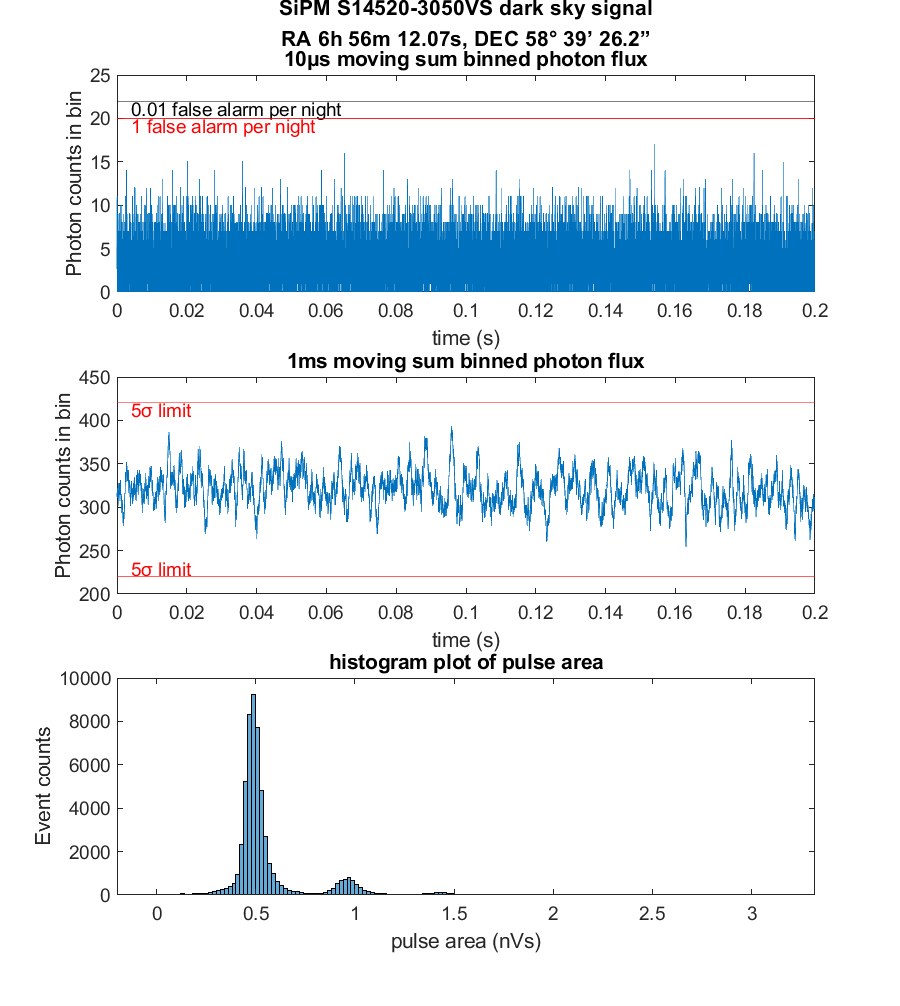}
    \captionof{figure}{SiPM S14520-3050HS observing area without star brighter than 18 mag (G-band)}
    \label{fig:s14520_empty}
\end{center}
\begin{center}
    \includegraphics[width=1.1\textwidth]{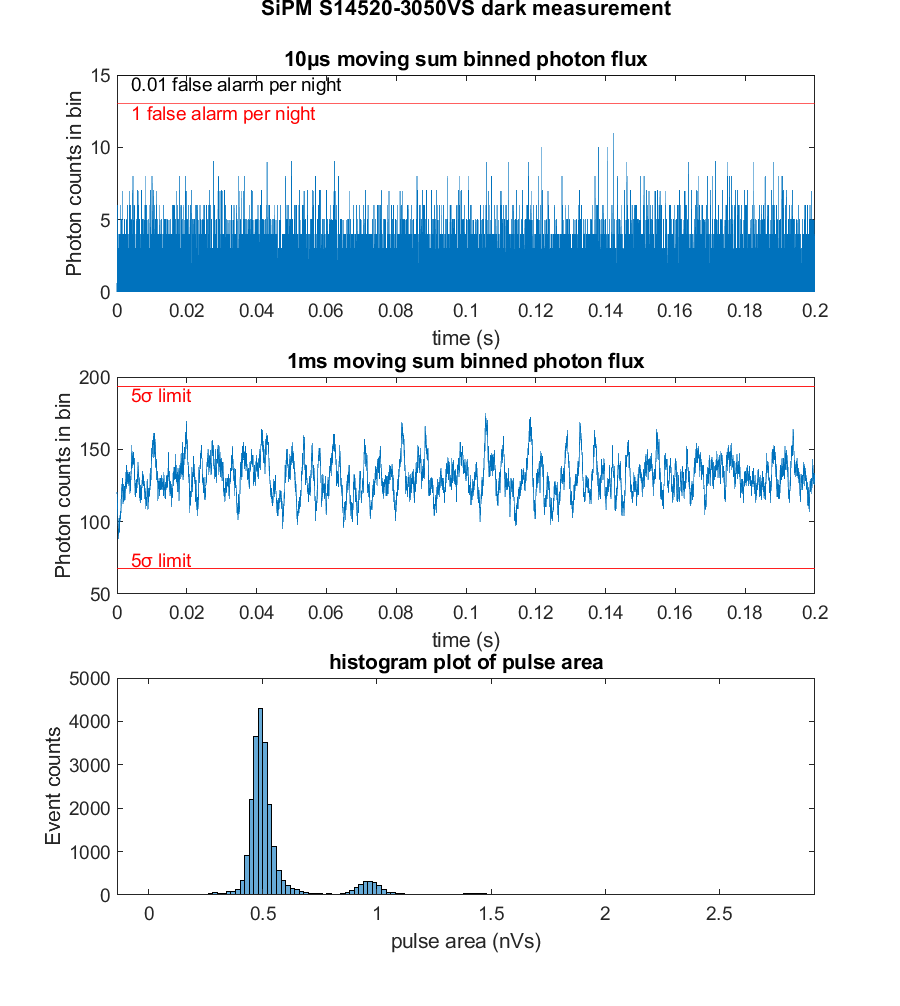}
    \captionof{figure}{SiPM S14520-3050HS dark measurement on site}
    \label{fig:s14520_dark}
\end{center}
\acknowledgments

We acknowledge support from RK MES grant AP05135753 through Nazarbayev University, Kazakhstan, and from the HKUST Jockey Club Institute for Advanced Study (IAS).

We would like to offer special thanks to the staff of the Assy-Turgen Observatory and the Fesenkov Astrophysical Institute, especially engineer Maxim Krugov, for their support, help and advice during the experiment.

We wish to acknowledge help on preparing the experiment by technicians Ulf Lampe and TK Cheng of the HKUST Physics Department.

This work has made use of data from the European Space Agency (ESA) mission
{\it Gaia} (\url{https://www.cosmos.esa.int/gaia}), processed by the {\it Gaia}
Data Processing and Analysis Consortium (DPAC,
\url{https://www.cosmos.esa.int/web/gaia/dpac/consortium}). 
Funding for the DPAC
has been provided by national institutions, in particular the institutions
participating in the {\it Gaia} Multilateral Agreement. 

\bibliography{main}   
\bibliographystyle{spiejour}   

\vspace{1ex}
Biographies of the authors are not available.

\listoffigures
\listoftables

\end{spacing}
\end{document}